\documentclass[runningheads]{llncs}
\usepackage{amsmath}
\usepackage{graphicx}
\usepackage{multirow}

\usepackage{colortbl}
\usepackage{hhline}
\usepackage{color}
\usepackage{tabularray}
\UseTblrLibrary{booktabs}

\begin{document}
\title{SAMDA: Leveraging SAM on Few-Shot Domain Adaptation for Electronic Microscopy Segmentation}
\titlerunning{Leveraging SAM on Few-Shot DA for EM Segmentation}
%
%
\author{Yiran Wang, Li Xiao*}
\authorrunning{PaperID 490}
%
\institute{School of Artificial Intelligence, Beijing University of Post and Telecommunication,Beijing 100876,China
\email{andrewxiao@bupt.edu.cn;andrew.lxiao@gmail.com}\\
\url{}}
\maketitle
\begin{abstract}
It has been shown that traditional deep learning methods for electronic microscopy segmentation usually suffer from low transferability when samples and annotations are limited, while large-scale vision foundation models are more robust when transferring between different domains but facing sub-optimal improvement under fine-tuning. In this work, we present a new few-shot domain adaptation framework \textbf{SAMDA}, which combines the Segment Anything Model(SAM) with nnUNet in the embedding space to achieve high transferability and accuracy. Specifically, we choose the Unet-based network as the "expert" component to learn segmentation features efficiently and design a SAM-based adaptation module as the "generic" component for domain transfer. By amalgamating the "generic" and "expert" components, we mitigate the modality imbalance in the complex pre-training knowledge inherent to large-scale Vision Foundation models and the challenge of transferability inherent to traditional neural networks. The effectiveness of our model is evaluated on two electron microscopic image datasets with different modalities for mitochondria segmentation, which improves the dice coefficient on the target domain by 6.7\%. Also, the SAM-based adaptor performs significantly better with \textbf{only a single annotated image} than the 10-shot domain adaptation on nnUNet. We further verify our model on four MRI datasets from different sources to prove its generalization ability.  

\end{abstract}
\section{Introduction}
\par Mitochondria segmentation from electron microscopic (EM) images is pivotal for understanding cellular processes in health care and medical research, given mitochondria's essential role in energy production within eukaryotic cells. This task is challenging due to mitochondria's variable size, shape, and distribution, complicating their detection and delineating their boundaries from surrounding structures. Accurate segmentation is crucial for insights into their structures and functions\cite{deepcontact}, driving advancements in mitochondrial segmentation techniques to the forefront of biomedical research with broad scientific implications. With the rapid development of deep learning techniques, various models with curated datasets are released for Mitochondria segmentation. However, a crucial issue exists: the feature distributions for the EM data vary with different machines and samples, and annotation for the EM data is time-consuming and expertise-demanding\cite{mitoem}. Therefore, there is an urgent need to develop an efficient and effective domain adaptation framework to improve model robustness and transferability.

\par The UNet model\cite{UNet}, characterized by its U-shaped structure, enables a dense fusion of shallow and deep features. With its subsequent network, including UNet++\cite{UNet++}, nnUNet\cite{nnUNet}, and MedNext\cite{MedNext}, the UNet-type network has gained significant success in medical image segmentation tasks. Meanwhile, Large-scale Vision Foundation Models such as SAM\cite{SAM}, SAM-MED\cite{SAMMed,sammed3d}, and MedSAM\cite{MedSAM} have shown superior performance in image segmentation tasks. The Segment Anything Model (SAM), with its Visual Transformer(ViT)\cite{vit} encoder rich in pre-training knowledge, demonstrates more stability and robustness when facing distribution change and domain shift of the data. In this work, we propose a novel protocol combining a SAM-based model with nnUNet, named \textbf{SAMDA}, which enables rapid transfer from the source domain to the target domain under few-shot conditions.

\par Domain-agnostic representation learning is widely adopted in domain adaptation, which minimizes the divergences between data representations in the source and target domains to achieve efficient domain adaptation. Traditionally, the Maximum Mean Discrepancy (MMD)\cite{mmd} is a dominant metric for measuring divergence between different sources. However, the MMD only measures the feature embedding distance but ignores the local discriminative features, which is unsuitable for EM segmentation tasks as it needs to identify a large amount of organelle features locally. Recently, adversarial training strategies have been developed rapidly, which provide an implicit way to learn and transfer local features and are widely applied in domain adaptation\cite{DA}.The
 \textbf{perceptual loss}\cite{loss},which is a commonly used loss function in style transfer(such as CycleGAN\cite{cycleGAN}) with adversarial training strategy\cite{dehazing}, has been proven to achieve stable and high-quality local feature transfer. Here, we adopted the perceptual loss into the Domain-agnostic representations learning framework for the EM data domain transfer.

\par Our proposed \textbf{SAMDA} consists of three steps of training:

\par Step 1: we concatenate embeddings of the SAM-based encoder and UNet encoder inherited from nnUNet to form a new segmentation model trained on the source domain;

\par Step 2: we freeze the UNet and adopt the SAM-based module as a "generic" model to perform domain adaptation, and propose a domain-agnostic representations learning strategy with perceptual loss; 

\par Step 3:we use the few-shot target sources to fine-tune the model further to improve the segmentation performance on the target domain.

\par In this way, we \textbf{utilize the rich pre-trained knowledge of the ViT encoder to better capture feature variances between different domains while retaining the benefit of efficient segmentation learning by nnUNet}. Our model is verified on two Electron Microscopy datasets: the EPFL(FIB-SEM) and Kasthuri++(ssSEM) datasets. Our SAM-based model demonstrate superior transferabiltiy compare to nnUNet, leading to a performance boosting by 6.7\% of Dice Coefficient. Our model is further verified on 4 MRI datasets from different sources. To demonstrate its generalizability, we test all the possible pairs of the MRI sources for domain transfer. All 12 experiments show that the SAM-based adaptor consistently improves the domain adaptation performance. 

\section{Methodology}
\par
\begin{figure}
    \centering
    \includegraphics[width=14cm]{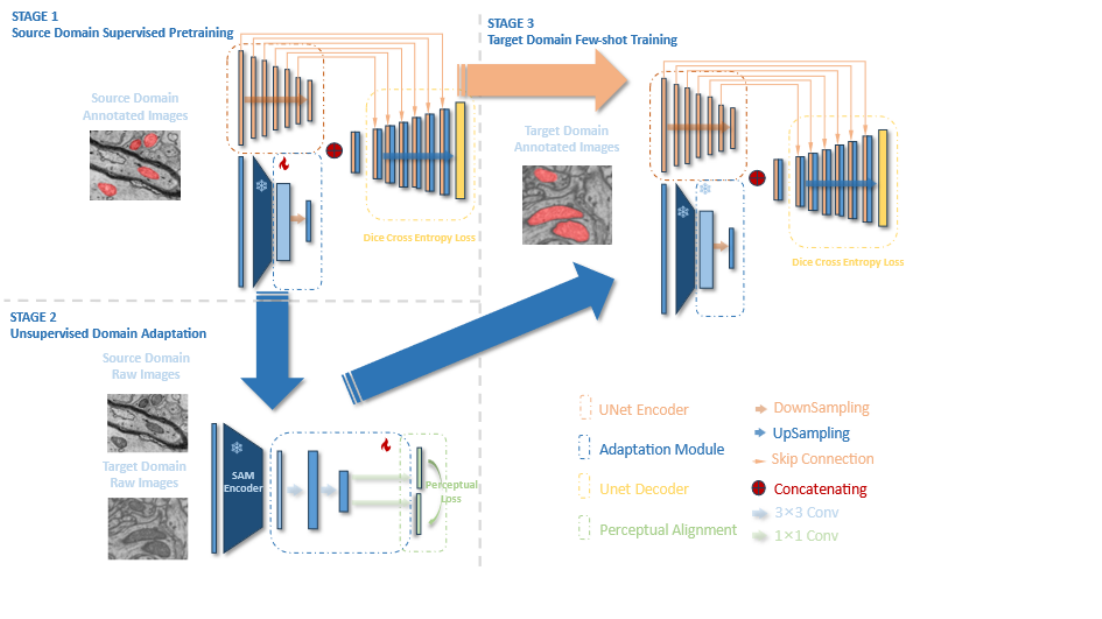}
    \caption{Overview of the 3-step Domain Adaptation Framework \textbf{SAMDA}. }
    \label{model}
\end{figure}

 Let's define a source domain dataset $D_S$, which is consisted by annotated and unannotated(raw) images denoted as $\left( S_{raw},S_{annotated} \right) \subset D_S$. The target domain dataset $D_T$ also contains annotated and raw images denoted as $\left( T_{raw},T_{annotated} \right) \subset D_T$. Assume $T_{annotated}$ only contains very few images thus there forms a few shot domain adaptation task. The whole protocol is depicted in Figure \ref{model}, which consists three stages: 1.the \textbf{source domain supervised training} step develops a semantic segmentation model $M_{pretrain}$ in a fully supervised manner based on the $S_{annotated}$ dataset; 2.the \textbf{unsupervised domain adaptation} step conduct unsupervised domain adaptation training based on the unannotated images $T_{raw}$ and $S_{raw}$, and obtains a domain-adapted model $M_{uda}$ based on $M_{pretrain}$; 3. the \textbf{target domain few-shot training} step optimizes $M_{uda}$ with the few-shot target source  $T_{annotated}$. Details are summarized below. 

\subsection{Model}
\subsubsection{nnUNet Backbone}
\par We choose this base model as the plug-and-play medical image segmentation framework nnUNet and adopt its \textbf{built-in 2D version of UNet as the segmentation backbone}. The Unet Encoder consists of eight stages $E_1,E_2,E_3,E_4,E_5,E_6,E_7,E_8$ and each stage containing max pooling and convolutional layers. The Unet Decoder contains seven stages: $D_1,D_2,D_3,D_4,D_5,D_6,D_7$, and each stage containing up pooling and convolutional layers. Meanwhile, skip connections are performed between $\left( E_i,D_i \right) \ i\in \left[ 1,7 \right]$.

\subsubsection{SAM-Based Adaptation Module}
\par To improve the model's robustness and transferability, we propose a SAM-based adaptation module and combine it with the UNet backbone. The module consists of two parts: the first part is the SAM pre-trained encoder, structured as a Vision Transformer(ViT); the second part is an embedding projector with three convolutional layers, whose output feature map will be concatenated with the feature embedding of the last layer of the UNet Encoder($E_8$) and then fed into the UNet Decoder together.

\subsection{Training Procedure}
\par The training procedure is consisted by three sequential steps.
\subsubsection{Source Domain Supervised Training}
In the first step, the source domain $S_{annotated}$ is taken as the input and processed by the UNet Encoder and SAM-Based Adaptation Module separately to obtain two feature maps $h_{UNet},h_{SAM}$. The two feature maps are then concatenated and connected to the UNet Decoder to form the segmentation model. We choose the loss function as the Dice Cross Entropy loss: 
\begin{equation}\label{eq1}
        L_{DCE}=L_{ce}+L_{dice}
\end{equation}
After training, the SAM-based adaptor will encode the source domain features consistent with Unet encoder.
\subsubsection{Unsupervised Domain Adaptation}
The second step will freeze the UNet and let the SAM-Based Adaptation Module $\phi_s$ stand-alone for domain adaptation. The SAM encoder receives paired unlabeled inputs from the source domain $S_{raw}$ and the target domain $T_{raw}$ and generates two feature maps $h_s,h_t$ of matching sizes. To enable this module to comprehend and integrate deep semantic information from both domains simultaneously, we design a perceptual loss which is composed of Feature Reconstruction Loss and Style Reconstruction Loss, formulated as:
\begin{equation}\label{eq2}
    \begin{split}
        L_{perceptual}&\left( s_{raw},t_{raw} \right) =L_{feat}\left( s_{raw},t_{raw} \right)+L_{style}\left( s_{raw},t_{raw} \right)\\ 
    \end{split}
\end{equation}
Here C, H, and W respectively represent the number of channels and height, width of the feature maps. The Feature Reconstruction Loss $L_{feat}$ minimizes the Euclidean distance between the source domain and target domain input features after being processed by the SAM encoder $\phi _s$.
\begin{equation}\label{eq3}
    \begin{split}
        L_{feat}\left( s_{raw},t_{raw} \right)=\frac{1}{CHW}\lVert \left. \phi _s\left( s_{raw} \right) -\phi _s\left( t_{raw} \right) \rVert_{2}^{2} \right.\\
    \end{split}
\end{equation}
To obtain the Style Reconstruction Loss $L_{style}$, we compute the Gram matrices $G\left(s\right)$ for the source and target inputs $s_{raw},t_{raw}$, which involves computing the pairwise inner products of two feature maps across all channels. The Gram matrix is considered as an important indicator representing the style of an image, which is a C×C matrix denoted as $G(s)=\{G_{ij}(s)|i,j \in [0, C-1]\}$. Here
\begin{equation}\label{eq5}
    \begin{split}
        G_{i,j}\left( s \right) =\frac{1}{CHW}\sum_{h=1}^H{\sum_{w=1}^W{\phi \left( s \right) _{h,w,c^i}}}\phi \left( s \right) _{h,w,c^j}
    \end{split}
\end{equation}
where $c^i,c^j$ represent different channels. The $L_{style}$ is then computed by calculating the Euclidean distance between the Gram matrices
\begin{equation}\label{eq4}
    \begin{split}
         L_{style}\left( s_{raw},t_{raw} \right) =\lVert \left. G\left( s_{raw} \right) -G\left( t_{raw} \right) \rVert _{2}^{2} \right. 
    \end{split}
\end{equation}
In this way, the SAM-based adaptor narrows the domain gap in the embedding space and achieves domain transfer unsupervisedly. 
\subsubsection{Target Domain Few-shot Training}
We further adopt the few-shot target source $T_{annotated}$ to train the segmentation model. During the few-shot training module, both UNet- and SAM-based encoders are finetuned to adapt better to the target source distribution. The loss function remains as the Dice Cross Entropy Loss.

\section{Experimental setup}
\subsection{Datasets}
Experiments are performed on two datasets with different electron microscopic modalities, the \textbf{Kasthuri++} and \textbf{EPFL} dataset. To further verify the generalizabilty of our model, We also performed domain adaptaion experiments on four hippocampus MRI datasets from different sources. 
\subsubsection{EM datasets} contain two datasets with different modalities:1.\textbf{Kasthuri++ dataset}\cite{Kasthuri} is consisted by annotated serial section electron microscopy (ssEM) images with an anisotropic resolution of 3×3×30 nm per voxel, The training set contains 85 images with size of 1463×1613 and the testing set contains 75 images with size of 1334×1553, respectively; 2.\textbf{EPFL dataset}\cite{EPFL} contains 165 annotated focused ion annotated beam scanning electron microscopy (FIB-SEM) images with an isotropic resolution of 5×5×5 nm per voxel and size of 1024*768.
\subsubsection{MRI Datasets} contain four MRI scans of the hippocampus datasets with annotations are adopted for domain adaptation: 1.\textbf{HarP dataset}\cite{HarP} contains 135 MRI scans, demented subjects with age from 60 to 90, 1.5T/3T GE scanners; 2.\textbf{Hammers dataset}\cite{Hammers} contains 30 MRI scans, healthy subjects with age from 20 to 54, 1.5T GE scanner; 3.\textbf{Oasis dataset}\cite{Oasis} contains 35 MRI scans, healthy subjects with age from 18 to 90, 1.5T GE scanner; 4.\textbf{LPBA40 dataset}\cite{LPBA40} contains 30 MRI scans, healthy subjects with aged from 20 to 54, 1.5T GE scanner.

\subsection{Networks and training settings}
\par Our models are implemented in PyTorch and accelerated by an NVIDIA A100 80G GPU. The parameters of the UNet module are inherited from nnUNet, and the parameters of the SAM ViT encoder are frozen following the typical fine-tuning operation. We adopt different pre-trained backbones for experiments for the SAM-based adaptation module, including MedSAM\cite{MedSAM}, SAM-MED2D\cite{SAMMed}, and Mobile SAM\cite{mobilesam}. During the first step, we conduct source domain supervised training for 100 epochs to ensure convergence, and we train the model 10 epochs during the second step and 50 epochs during the third step. The initial learning rates during the first and second steps are set as $1 \times 10^{-2}$ and set as $1 \times 10^{-3}$ during the third step. The weight decay of the learning rate is set to $3 \times 10^{-5}$ during the three steps.

\section{Result and Discussion}
\par Firstly, to validate the effectiveness of our domain adaptation protocol in unsupervised and few-shot scenarios, we choose the domain transfer from the Kasthuri++ dataset to the EPFL dataset as a typical scenario and establish a six-group domain adaptation task with 1, 2, 4, 8, 10, and 20 annotated samples, respectively (number of target samples used in step 3). We randomly pick 100 images from EPFL dataset as the target domain testing set, and the remaining images form the target domain training set. Concurrently, to discuss the effectiveness of the pre-trained knowledge in the ViT Encoder, we selected three versions of SAM, the MedSAM, SAM-MED, and MobileSAM, for comparison. 
\par
\begin{figure}[htb]
    \centering
    \includegraphics[width=8cm]{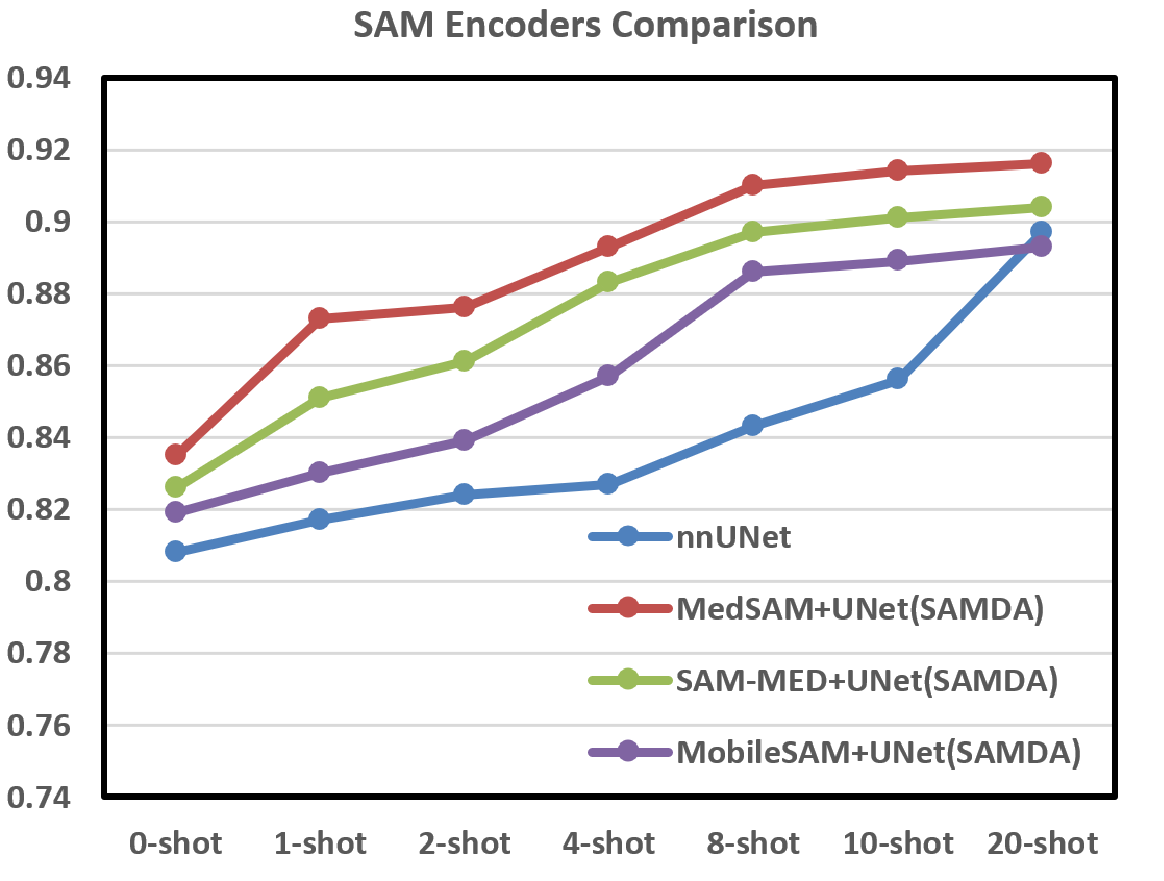}
    \caption{A line graph presentation of the segmentation results of Table \ref{table1} with increased number of annotated samples. The Dice coefficient is taken as the evaluation metric.}
    \label{fig1}
\end{figure}

 As shown in Table \ref{table1} and Figure \ref{fig1}, all three SAM-based encoders demonstrate superior transferability than nnUNet under all the few-shot conditions. The MedSAM version, pre-trained with rich medical knowledge, exhibits the best performance. It is interesting to notice that with the assistance of the MedSAM encoder, our model significantly outperforms the 10-shot domain adaptation on nnUNet with \textbf{only a single annotated image}.
\begin{table}[htb]
    \centering
    \includegraphics[width=\textwidth]{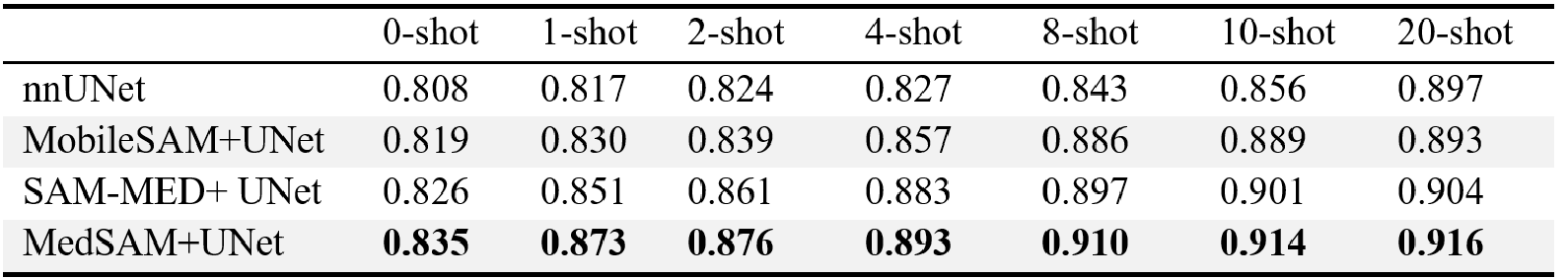}
    \caption{Domain adaptation performances for three versions of SAM encoder and nnUNet under various scenarios with different number of target samples. The best result is marked with bolding.}
    \label{table1}
\end{table}

\par To further validate the generalization ability and effectiveness of our protocol, we extend our few-shot domain adaptation tasks on four MRI datasets. For EM data, we choose one dataset as the source and another as the target. We keep the 75 testing images as the testing set for the Kasthuri++ dataset and randomly pick 100 images to form the testing set for the EPFL dataset. For MRI data, we choose one dataset as the source and select one of the other as the target, and randomly selected 1/5 of the annotated images from target dataset to form the testing set. To standardize our tasks, each round we select 20 pairs of raw images for step 2 and 10 annotated target images for step 3. The Dice Coefficient(DC) is adopted as the evaluation metric. The results are summarized in Table \ref{table1}. We report four sets of results:1.nnUNet only(noDA):we test the nnUNet on the target source directly without domain adaptation; 2.nnUNet Only(withDA): we take UNet as the adaptor for domain adaptation;3.SAMDA(0 shot): we take MedSAM as the adaptor encoder but only perform unsupervised domain adaptation(without step 3);4.SAMDA(10 shots): we take MedSAM as the adaptor encoder since this version performed best in the few-shot experiment,and perform the whole domain adaptation process.

\begin{figure}[htb]
    \includegraphics[width=\textwidth]{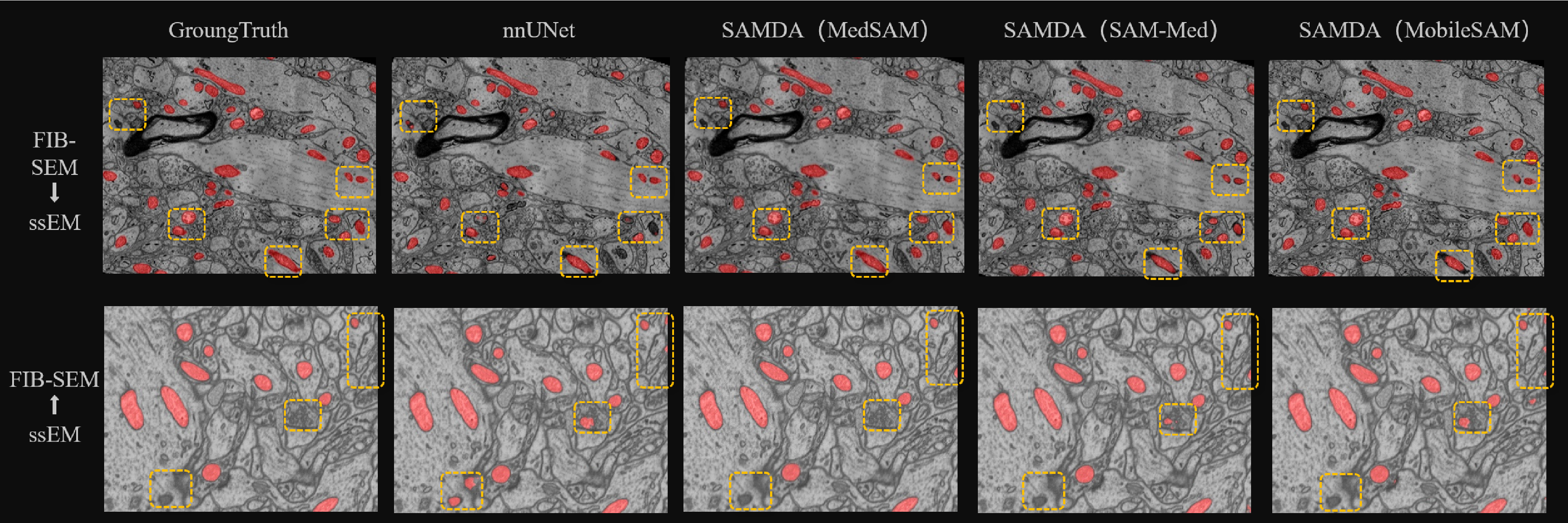}
    \caption{Visualization of domain adaptation results between FIB-SEM and ssEM data is presented.}
    \label{fig3}
\end{figure}

The results show that for all the 14 groups of experiments, the domain adaptation process consistently improve the performance. There is an average increase of DC by 17\% (from 0.643 to 0.753) wiht the unsupervised domain adaptation and a continued rise of DC by 7.4\%(from 0.753 to 0.809) with the few-shot training. Meanwhile, the SAM encoder demonstrates its excellent transferability, with an average increase of DC by 2.7\%(from 0.788 to 0.809). Especially for Mitochondria segmentation when transferring from Kasthuri++ dataset to EPFL dataset, SAM outperforms nnUNet by 6.7\%.
\begin{table}
    \centering
    \includegraphics[width=\textwidth]{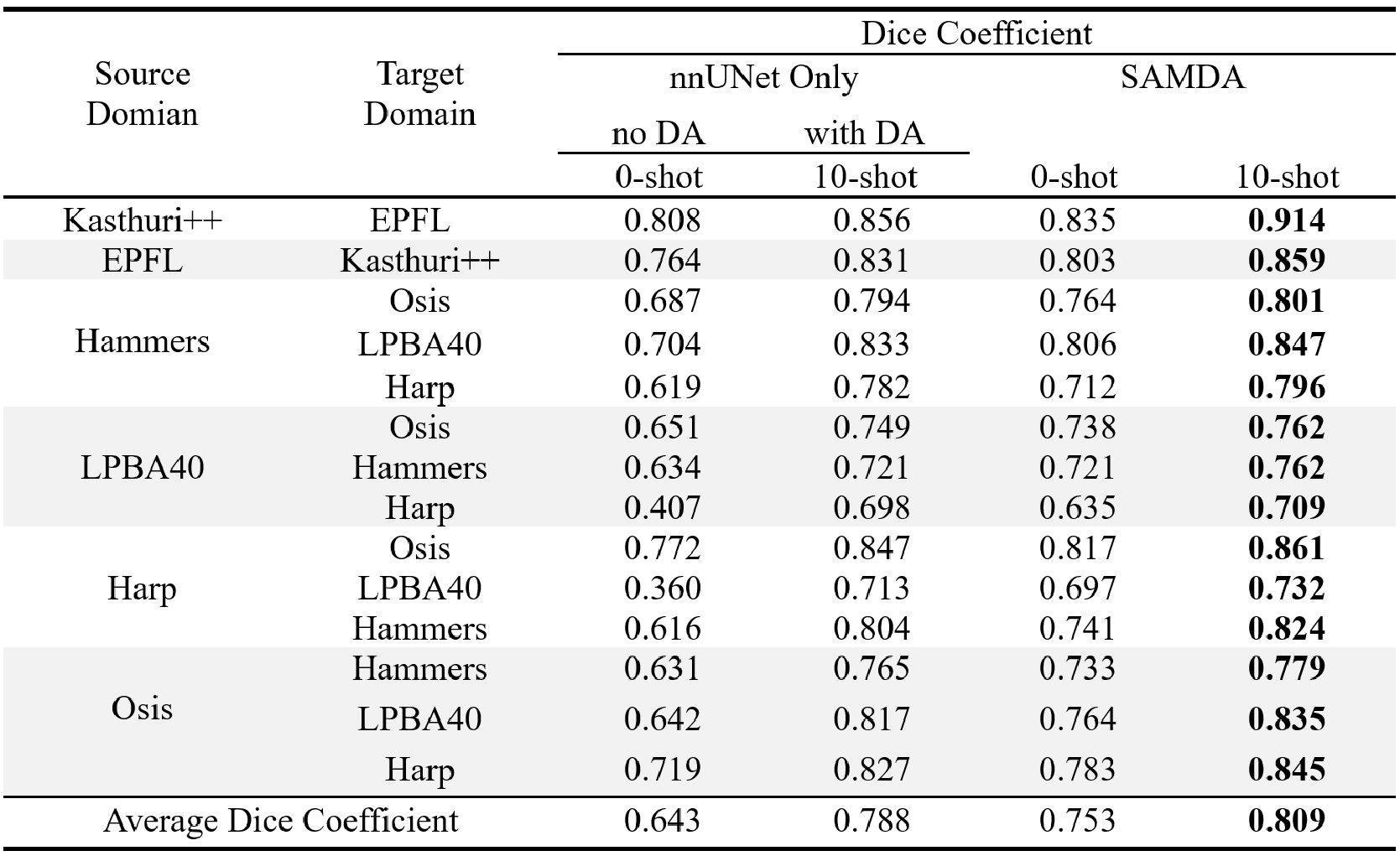}
    \caption{Quantitative comparisons of segmentation performances on 14 rounds of domain adaptation tasks for EM and MRI datasets. The Dice Coefficient is taken as the evaluation metric. The best results are marked with bolding.}
    \label{table2}
\end{table}
In Figue \ref{fig3}, we also present the visualization results of mitochondria segmentation in few-shot scenario. With assistance from the SAM-based adaptor, our model achieves better segmentation performance on the target domain, especially when segmenting mitochondria distributed within complex backgrounds.

\section{Conclusion}
\par This work introduces a novel approach for few-shot domain adaptation by effectively integrating the Segment Anything Model (SAM) with the nnUNet framework on the embedding space to achieve high transfer ability while retaining accuracy and efficiency. The performance of our proposed model was rigorously tested on various datasets, including two EM datasets of mitochondria segmentation and four MRI datasets of hippocampus segmentation, demonstrating a notable performance boosting on domain adaptation, especially on few-shot conditions. In the future, we will generalize this domain adaptation approach with other large-scale Vision Foundation models and tasks, for example, image registration, report generation,et.al.

%
%
%
\bibliographystyle{splncs04}
\bibliography{mybibliography}
\end{document}